\documentstyle[twocolumn,psfig,aps]{revtex}
\begin{document}
\draft
\title{Surface Phonons and Other Localized Excitations}
\author{Philip B. Allen, Seth Aubin}
\address{Department of Physics and Astronomy, State University of New York,
Stony Brook, New York 11794-3800}
\author{R. B. Doak}
\address{Department of Physics and Astronomy, Arizona State University,
Tempe, Arizona 85287-1504}
\date{\today}
\maketitle
\begin{abstract}
The diatomic linear chain of masses coupled by harmonic springs
is a textboook model for vibrational normal modes (phonons) in
crystals.  In addition to propagating acoustic and optic branches,
this model is
known to support a ``gap mode'' localized at the surface, provided
the atom at the surface has light rather than heavy mass.
An elementary argument is given which explains this mode and
provides values for the
frequency and localization length.  By reinterpreting this mode
in different ways, we obtain the frequency and localization
lengths for three other interesting modes: (1) the surface vibrational
mode of a light mass impurity at the surface of a monatomic chain; (2) the
localized vibrational mode of a stacking fault in a diatomic chain;
and (3) the localized vibrational mode of a light mass impurity 
in a monatomic chain.
\end{abstract}

\section{introduction}

Unlike molecules which
have discrete vibrational frequencies, crystals have a continuous
spectrum of vibrations which can propagate
as travelling waves \cite{Kittel,Ashcroft,Ibach}.
This fact causes crystals to be much better
heat conductors than glasses or liquids.
Sometimes the spectrum is interrupted by gaps where no
propagating normal modes occur.  
Other interesting behavior happens at frequencies inside the gap, such
as localized (non-propagating) normal modes associated with defects
and surfaces.  The text by Ziman
\cite{Ziman} has a good discussion.  A visualization
of surface modes on the (100) surface of Cu is on the website of Ch. W\"oll,
Ruhr-Universit\"at Bochum \cite{Woll}.
The present paper shows
how this happens for some particular cases of one-dimensional
crystals, or linear chains of atoms.  Our treatment uses only classical
mechanics, and gives properties (frequency and displacement
pattern) rigorously by pictorial arguments with no higher algebra.

The surface phonon provides the simplest
example of wave localization, an effect
which occurs in many branches of physics.
Analogous phenomena are found in the quantum treatment of electrons
in single-particle approximation \cite{Kittel,Ashcroft},
and in the new field of ``photonic band-gap systems''  \cite{Joann}.
This paper reports a simple way of understanding the surface
phonon on the diatomic linear chain.  The model is then
extended and reinterpreted to give simple explanations
of some other localized modes.

\section{diatomic molecule}

A diatomic molecule has a single vibrational ``normal mode.''
Even though the restoring force of atom 1 on atom 2 has in reality
a complicated quantum-mechanical origin,
for small displacements away from equilibrium, it can always
be well approximated by a spring obeying Hooke's law with
a spring constant $K$.  Using standard physics of the
two-body problem \cite{French}, if the two
atoms have masses $M_H$ and M$_L$ ($H$ and $L$ are for
heavy and light), the squared oscillation frequency 
$\omega^2$ is $K/M_{\rm red}$ where $M_{\rm red}$ is the ``reduced mass'' 
$M_L M_H/(M_L+M_H)$.  

\section{perfect infinite chains}

A crystalline solid is a
very large molecule, with a continuous spectrum (or band) of vibrational
frequencies.  Solids can also be modelled by masses
connected to each other by springs.  A one-dimensional chain of
masses is often studied, not because it is
found in nature, but because the mathematics is
simple and can be generalized to more realistic three-dimensional
arrangements.  For a large enough collection of atoms, most of
the vibrational normal modes are classified as ``bulk'' normal
modes, which means they are essentially identical to those of a hypothetical
infinite sample with no boundaries.  Each ``bulk'' normal mode has
a pattern of atomic displacements which extends throughout the 
system.  Similar to the normal modes of a vibrating string,
these are sine and cosine standing waves.  Alternately, 
one can use linear combinations of sines and cosines to give
an equivalent basis of left and right-going
travelling waves.  For the case of all masses equal to $M_0$, 
the $\ell$'th atom (located at $R_{\ell}=\ell a$) 
has a displacement $A\sin(kR_{\ell}-\omega_k t)$ 
in a right-going travelling wave.
The corresponding squared frequency is $(4K/M_0)\sin^2(ka/2)$.  
There are as many such solutions ($N$) as there are atoms in the
chain, namely solutions for each
$k$ in the range $(-\pi/a,\pi/a)$.  This is derived in many texts
\cite{Kittel,Ashcroft,Ibach,Goldstein}.  For $N \rightarrow \infty$ 
the spectrum is continuous between
the minimum squared frequency of zero and the maximum of 
$\omega_{\rm MAX}^2=4K/M_0$.
A particularly original discussion is given by Martinez \cite{Martinez}.

\begin{figure}
\centerline{
\psfig{figure=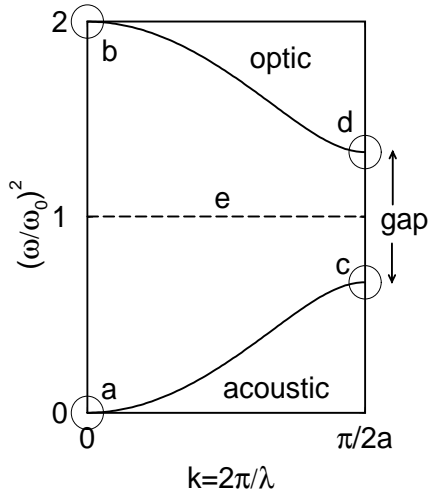,height=2.7in,width=3.0in,angle=0}}
\caption{Dispersion of squared frequency (in units $\omega_0^2=K/M$)
versus wavevector $k$ for the diatomic chain.  
The gap is proportional to $(M_H-M_L)/(M_H+M_L)$,
and is drawn for the case $M_H=2M_L$.  The distance between atoms is $a$.
The dashed line shows the position of the surface mode.}
\label{disp}
\end{figure}

The vibrational spectrum of a real material sometimes has a 
gap, an interval of frequencies where there are no travelling
wave solutions.  A simple
model illustrating this is the ``diatomic chain,'' 
an infinite chain of alternating masses $M_L$, $M_H$.
The algebra, which is more complicated than the monatomic
chain, is also given in texts \cite{Kittel,Ibach}. 
The dispersion curve for $\omega_k^2$ is given in Fig. \ref{disp}.
There are now two ``branches,''  labelled
acoustic and optic, and a gap.  Exactly in the middle of the gap,
the surface may induce a {\bf localized} vibrational normal mode,
with amplitude which falls exponentially ($\propto \exp(-R/\xi)$)
with distance $R$ into the bulk.

Before discussing this, we sharpen our understanding with a
quantitative interpretation of the four special bulk modes
indicated by circles in Fig. \ref{disp}.  The frequencies of these
special modes can be understood without the algebra needed to find
the frequencies of the modes at general $k$-vectors.

\section{special bulk modes}

\begin{figure}
\centerline{
\psfig{figure=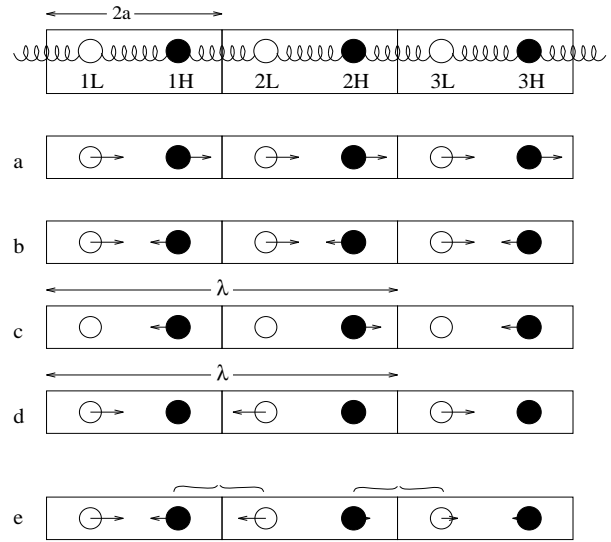,height=2.8in,width=3.1in,angle=0}}
\smallskip
\caption{The diatomic chain, the special bulk modes
\protect${\bf a, \ b}$ with wavelength \protect$\lambda=\infty$ and
wavevector $k=0$, the special bulk modes \protect${\bf c, \ d}$
with wavelength \protect$\lambda=4a$ and $k=\pi/2a$ (all shown circled
in Fig. \protect\ref{disp}), and the gap mode \protect${\bf e}$ 
confined at the
surface when a light atom terminates the chain.  In pictures 
\protect${\bf a-d}$, the chain is infinite;  
by contrast, in picture \protect${\bf e}$, the light atom at the left
is the surface atom, with no spring acting to its left.  The brackets
indicate pairs of atoms which move without altering the length of
the bond between them.}
\label{modes}
\end{figure}

The four special modes circled in Fig. \ref{disp}
have the simple vibrational patterns shown in Fig. \ref{modes}.  
First, why are these patterns ``normal modes''?  
If we take as initial conditions, the 
velocities of all atoms to be zero and the positions to be 
as shown in the figures, then Newton's laws have simple,
and perhaps even obvious solutions: the pattern is preserved,
and oscillates in time as $\cos(\omega t)$ for some special
choice of $\omega$.  This is the {\bf definition} of a normal mode.
Second, what is the corresponding frequency of oscillation?
This can be answered by careful consideration of forces and masses.

\begin{description}
\item[{\bf mode a:}]
This is the simplest mode with all atoms
having the same displacement.  This has infinite wavelength 
(zero wavevector), no stretch
of any spring, and therefore zero restoring force and $\omega=0$.

\item[{\bf mode b:}]
This has oppositely directed displacements for adjacent atoms.
Each unit cell of the crystal has the same displacment pattern.
Therefore the wavelength is infinite and the wavevector is zero.  
The displacements in mode ${\bf b}$ are such that $u_L$ (the displacement
of the light atom) is proportional to $M_H$, and similarly $u_H$ 
is proportional to $M_L$.  Thus the center of mass of each unit cell is fixed.
The mode is almost the same as in a diatomic molecule, except
each atom has two springs attached, one stretched and the other compressed
by the same amount.  Therefore, when released from rest, each pair of
atoms oscillates with fixed center of mass but with twice the
restoring force of an isolated diatomic molecule, {\sl i.e.} 
$\omega^2=2K/M_{\rm red}$.  This is the highest frequency
normal mode in the spectrum.  

\item[{\bf mode c:}]
This has light atoms 
stationary and heavy atoms moving in an alternating pattern.
The light atoms feel equal and opposite forces which cancel, while
the heavy atoms feel repulsive and attractive forces which add.
This pattern also oscillates in time, 
with squared frequency $\omega^2=2K/M_H$.

\item[{\bf mode d:}]
This is the same as mode {\bf c} 
except heavy and light atoms are interchanged, making the
squared frequency equal to $2K/M_L$.
Modes {\bf c} and {\bf d} have wavelength $4a$ and wavevector $\pi/2a$.
All other normal modes of the infinite crystal are more complicated and have
frequencies which lie on smooth curves connecting these four modes.
\end{description}

\section{surface mode in the gap}

Modes which are confined to the surface
region normally must have frequencies which lie outside the ``bulk'' bands.
Discussions of such modes are given in texts on surface physics
\cite{Zangwill,Desjon,Cottam,Luth,Davison}
and measurements are cataloged by Kress
and de Wette \cite{Kress}.
We have discovered a very simple explanation of the fact \cite{Wallis}
that a ``gap mode" confined to the surface occurs
in the diatomic chain if the endmost atom is a light atom.

Consider {\bf mode $e$}, which
like mode {\bf $b$} has pairs of atoms vibrating with fixed center
of mass.  However, adjacent {\bf pairs} vibrate in such a way that the
connecting spring is not stretched.  Thus each pair experiences no
force from any other atom and is decoupled from
the rest of the chain.  The resulting decoupled pairs oscillate with 
$\omega^2=K/M_{\rm red}$ as for isolated diatomic molecules.
Since all pairs have the same frequency, this is a stable normal mode.
The frequency lies exactly in the middle of the gap of
the squared frequency spectrum ($K/M_{\rm red}=(1/2)(2K/M_L+2K/M_H)$).
In order to be decoupled, the heavy atom of a given pair, and the
adjacent light atom of the next pair deeper into the bulk, must have
the same displacement, smaller by $M_L/M_H$ (and with opposite sign)
than the displacement of the previous light atom closer to the surface,
in order to conserve center of mass position.
Since adjacent pairs have displacement ratios
$-M_L/M_H$,  the $n$-th pair has amplitude proportional to 
$(-M_L/M_H)^n=(-1)^n \exp(-n\ln(M_H/M_L))$.
This is an exponential decay $\exp(-2na/\xi)$
with decay length $\xi=2a/\ln(M_H/M_L)$.  If the surface atom
had been a heavy atom, this mode would have been exponentially
growing rather than decaying, which is not allowed for a normal
mode.  Mode {\bf e} was first found by Wallis \cite{Wallis}
in an elegant calculation
of the spectrum of finite chains.  Our simple argument is not (to
our knowledge) in the literature.
A ``standard" derivation is given in the text by Cottam and Tilley 
\cite{Cottam}.

\section{surface modes of three-dimensional crystals}

\begin{figure}
\centerline{
\psfig{figure=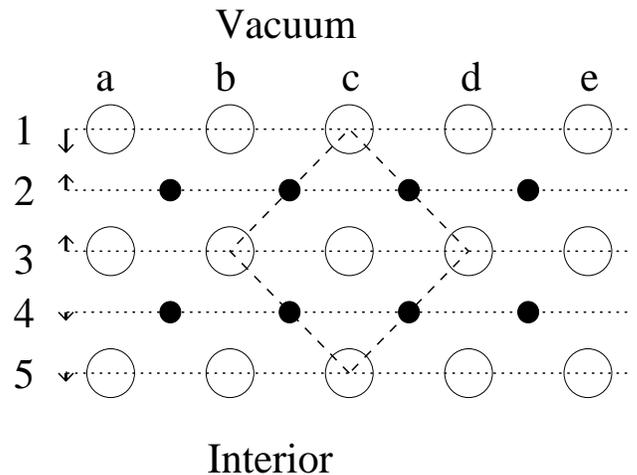,height=2.7in,width=3.5in,angle=0}}
\caption{``Polar'' surface of a diatomic crystal in two dimensions.
This is a two-dimensional analog of the ``(111)'' surface of crystals
with the ``rocksalt'' (NaCl) structure.  A square conventional unit
cell is shown as dashed lines parallel to conventional $\hat{x}$
and $\hat{y}$ axes.  There are two possible terminations
of such crystals, ones which expose planes of light atoms as shown
in the figure by open circles, and ones which expose planes of heavy atoms, 
shown as filled circles.  Since the heavy and light atoms have opposite
ionic charges, such surfaces have a surface dipole and are called polar. }
\label{bulk}
\end{figure}

Mode {\bf e} is directly related to a {\bf branch} of surface normal
modes of  higher-dimensional diatomic crystals.  
A two-dimensional version is shown in
Fig. \ref{bulk}.  Various types of surfaces are possible for
such crystals.  If cut perpendicular to a conventional  $\hat{x}$
or $\hat{y}$ axis shown in the figure by dashed lines, the surface
contains equal numbers of $A$ and $B$ ions, and is referred to as
``non-polar.''  By contrast, the surface shown is a ``polar surface''
with a layer of light atoms exposed and
layers of heavy and light atoms alternating underneath.  There is a
vibrational normal mode in which each
{\bf layer} oscillates perpendicular to the surface (as indicated
by arrows) and which is localized at the surface.  Of course, in
real crystals the forces extend beyond first neighbors, so the 
displacement ratio $(-M_L/M_H)$ may not be exactly obeyed and the
squared frequency may not lie exactly at mid-gap, but the actual
behavior will mimic reasonably well the idealized one-dimensional
example of the previous section.  

There is actually not just one mode of this type, but a
{\bf branch} of such modes, with displacement patterns sinusoidally modulated
along the surface.  The one depicted in  Fig. \ref{bulk} has
the surface atoms ``a'', ``b'', ``c'', all moving in phase, corresponding
to an infinite wavelength, or zero wavevector, parallel to the surface.
The other extreme case of modulation is when atoms along the surface
are completely out of phase; when atom
``a'' moves down, atom ``b'' moves up, and
so forth, corresponding to a wavelength $\lambda=2\sqrt{2}a$ in the
plane of the surface.  Thus we anticipate a branch of surface excitations
with wavevectors lying in the plane of the surface.  
In order for such a mode to be exponentially
localized in the surface region, the frequency of oscillation must lie
in a gap where there are no corresponding bulk normal modes with
the same components of wavevector in the plane of the surface.
A gap is almost certain to occur for the case of zero wavevector, but
at increasing wavevectors the gap may disappear, and the mode ceases
to be localized near the surface.

Dimension two or three also opens new possibilities less directly
related to one-dimensional models, such as surface normal modes
with displacements in the plane of the surface.  Many branches
of surface normal modes have been seen experimentally by scattering
experiments.  Unfortunately we have not been able to locate in the
literature any observation of the simple mode illustrated in 
Fig. \ref{bulk}.  This is perhaps because polar surfaces are 
relatively unstable and hard to create and work with.

\section{impurity atom on the surface}
Another known result is that a surface mode appears above the
bulk frequency spectrum for a monatomic
chain, provided the atom on the surface is lighter than the rest by
at least a factor of two.  This can be proven by a reinterpretation
of the previous construction.  For mode ${\bf e}$ in Fig. \ref{modes},
let the two atoms connected by the unstretched spring be reinterpreted
as a single atom of mass $M_0=M_H + M_L$.  Then the model has
new interior atoms all with mass $M_0$, but a surface impurity atom with
mass $M_{\rm imp}=M_L < 0.5 M_0$.  The surface mode ${\bf e}$ still solves
Newton's laws with $\omega_S^2=K/M_{\rm red}$ and 
$M_{\rm red}=M_{L} M_H/(M_{L} + M_H)$.
In terms of the new variables $M_{\rm imp}$ and $M_0$
the reduced mass $M_{\rm red}$ is 
$M=M_{\rm imp} (M_0 - M_{\rm imp})/M_0$.  
The frequency $\omega_S^2$
lies above the top of the bulk band ($\omega_{\rm MAX}^2=4K/M_0$) 
if $M_0 > 2M_{\rm imp}$, and merges into the bulk band for
$M_0 \leq 2M_{\rm imp}$.  This result seems also to have
been first discovered by Wallis \cite{Wallis1}.  A ``standard'' proof of this
result is in the book by Desjonqu\`eres and Spanjaard \cite{Desjon}.

\section{localized gap mode of a stacking fault}

\begin{figure}
\centerline{
\psfig{figure=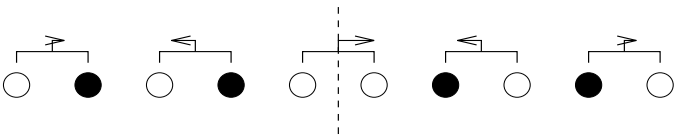,height=0.7in,width=3.5in,angle=0}}
\caption{Localized normal mode of vibration of the diatomic
chain with a stacking fault, obtained from mode \protect{\bf e} of
Fig. \protect\ref{modes} by reflecting the lattice and assigning
\protect$u(-\ell)=u(\ell)$}
\label{stacking}
\end{figure}

The gap mode {\bf e} of Fig. \ref{modes} generates a corresponding
mode of a defective bulk crystal, shown in Fig. \ref{stacking}.
This mode decays exponentially in both directions away from the
center of symmetry.  This center lies in the middle of a
``stacking fault'' where two light-mass atoms have been
put adjacent to each other.  It is a one-dimensional version
of a planar defect which occurs in real three-dimensional crystals.
The quantum-mechanical force between two light-mass atoms differs
from the force which binds the atoms of unlike mass.  Therefore,
we must expect that the separation $a^{\prime}$ of the light-mass
atoms will differ from the equilibrium separation $a$ of unlike atoms,
and that the force constant $K^{\prime}$ between these atoms
will differ from the constant $K$ occuring elsewhere.  Notice that
for the special displacement pattern of Fig. \ref{stacking}, there
is no force between the adjacent light atoms, so the
values of $a^{\prime}$ and $K^{\prime}$ are irrelevant; the squared
frequency of the normal mode is exactly the same as the surface
mode {\bf e} of Fig. \ref{modes}, and is pinned at midgap.

The stacking fault is a simple example of a ``topological
defect,'' that is, a defect which cannot be transformed away
by any local change.  As far as we know, the mid-gap normal
mode of vibration found here for the stacking fault has not
previously been discussed in the literature.  However, a close
analog is the ``topological soliton'' found at mid-gap in the
electronic spectrum of the ``Su-Schrieffer-Heeger'' model \cite{Heeger}
for polyacetylene with a topological defect in the pattern
of dimerization of carbon-carbon bonds along the chain.

\section{localized vibration of a light mass impurity in a monatomic chain} 

Suppose an impurity of mass $M_{\rm imp}<M_0$ is substituted into a monatomic
chain of mass $M_0$ with no change in force constants.  
Define the fractional mass deficit to be
$\epsilon=(M_0-M_{\rm imp})/M_0 >0$.  It is known that this system
supports a localized mode whose frequency ``splits off'' above the
frequency $\omega_{\rm MAX}$ of the uppermost bulk mode.  Specifically,
the mode has squared frequency $\omega_{\rm MAX}^2 /(1-\epsilon^2)$
and is localized around the impurity with localization length
$a/\ln((1+\epsilon)/(1-\epsilon))$.  The earliest presentation of
this mode known to us is by Montroll and Potts \cite{Montroll}.
The topic of localized modes in solids had been given a systematic
formulation in three earlier papers by Lifshits, available only in
Russian \cite{Lifshits}.  A textbook derivation is given by
Mih\'aly and Martin \cite{Mihaly}, and a nice qualitative discussion
is given by Harrison \cite{Harrison}.

These results follow rigorously by reinterpretation of Fig. \ref{stacking}.
Simply regard each pair of co-moving atoms as a single atom whose
mass is the sum of the two shown in the figure.  Thus $M_0$
is $M_H+M_L$, $M_{\rm imp}$ is $2M_L$, and the new lattice constant
$a$ is twice the previous distance $a$.  When the impurity mass is
heavier than the host mass, there is no longer a split-off bound
state, but instead a ``resonance'' within the bulk band.

In three-dimensional crystals the occurrence of a vibrational bound state
requires a minimum mass deficit $\epsilon$ which is model-dependent,
whereas our 1-d example has a bound state for arbitrarily small
mass deficit.  This is a classical discrete-system analog of
the continuum quantum-mechanical theorem that an attractive well
always has a bound state in a 1-d one-electron problem
(and also in 2-d) but requires a
critical well-depth in 3-d \cite{Landau}.  For the impurity
on the surface, however, we saw that even in 1d there is a critical mass
deficit of 1/2.   The quantum analog is that if the well
is at the edge of a 1-d half space (the other half of space
is impenetrable because of an infinite potential), then there
is a critical well-depth, equal to the well-depth at which the
second bound state appears for the symmetric well in the full
1-d space.

\section{summary}
Two simple surface phonons and two simple bound defect modes
in one-dimensional lattices have been quantitatively explained by
pictorial construction and elementary physics of the two-body
problem.  This is certainly not a complete catalog of interesting
localized modes, but we think that
these modes can serve as useful pedagogical models for phenomena in
several branches in physics.  

\acknowledgements
We thank A. A. Maradudin and N. Stojic for help.
This work was supported in part by NSF grant no. DMR-9725037.

\end{document}